\documentclass{llncs}
\usepackage{makeidx}  
\usepackage{amsmath}
\usepackage{amsfonts}
\usepackage{cite}
\usepackage{wrapfig}
\usepackage{graphicx}
\usepackage{color}
\usepackage{authblk}
\graphicspath{ {figures/} }

\begin{document}

\frontmatter          
\pagestyle{headings}  
\addtocmark{} 

\mainmatter              
%

\title{\large Stopping criteria for boosting automatic experimental design using real-time fMRI with Bayesian optimization} 

\titlerunning{}  
%

\author{\footnotesize $^{*}$Romy Lorenz$^{1,2}$, $^{*}$Ricardo P Monti$^{3}$, Ines R Violante$^1$, Aldo A Faisal$^2$, Christoforos Anagnostopoulos$^3$,
Robert Leech$^1$ and Giovanni Montana$^{3,4}$\\
{
$^1$Department of Medicine, Imperial College London, W12 0NN, UK \\
$^2$Department of Bioengineering, Imperial College London, SW7 2AZ, UK\\
$^3$Department of Mathematics, Imperial College London, SW7 2AZ, UK\\
$^4$Department of Biomedical Engineering, King's College London, SE1 7EH, UK\\
\emph{$^*$These authors contributed equally to this work.}}}

%
%
\tocauthor{}

\institute{}

\maketitle              

\begin{abstract}
Bayesian optimization has been proposed as a practical and efficient 
tool through which to tune parameters in many difficult settings.
Recently, such techniques have been combined with real-time fMRI to propose a 
novel framework which turns on its head the conventional functional neuroimaging approach.
This closed-loop method automatically designs the optimal experiment to evoke a desired target brain pattern.
One of the challenges associated with extending such methods to 
real-time brain imaging is the need for adequate stopping criteria, an aspect of Bayesian optimization which 
has received limited attention. In light of high scanning costs and limited attentional capacities of subjects
an accurate and reliable stopping criteria is essential. 
In order to address this issue we propose and empirically study the 
performance of two stopping criteria. 

\end{abstract}

\section{Introduction}

The central question addressed by functional neuroimaging is to elucidate the inter-relation between cognition and the brain. In a typical functional magnetic resonance imaging (fMRI) experiment, a neuroscientist selects a task that is associated with a particular cognitive process, and studies the brain regions that elicit a response \cite{Amaro}. 
However, it is well documented that the same brain region can be activated by inherently different tasks (e.g., the 
superior temporal sulcus has been termed the ``chameleon of the brain'' \cite{chameleon}), thus undermining the standard approach.
To date, there is no approach available that systematically investigates how many different tasks activate the same brain region or network of regions. 
Gaining insights into the complex relationship between the brain and cognition though
requires a more thorough investigation of this mapping. 

In order to address this concern, the \emph{Automatic Neuroscientist} \cite{Lorenz} has recently been proposed. This framework turns the conventional fMRI approach on its head by using real-time fMRI \cite{Weiskopf} in combination with a state-of-the-art Bayesian optimization technique \cite{BayesOptTutorial,Snoek}  to systematically unveil the relationship between different tasks and their neural responses within an individual. In a nutshell, the framework starts with a target brain pattern and automatically finds a set of tasks that maximally activates it. 
This is done using a brain-machine interface: the {subject}'s brain pattern is analyzed in real-time in response to the current task. Based on this, the Bayesian optimization algorithm proposes a task that will be presented to the subject in the next iteration. This closed-loop cycle continues until the optimal task is found.



While the \emph{Automatic Neuroscientist} was 
validated in a proof-of-principal study involving five subjects \cite{Lorenz},
there remains several open questions. In particular, in this work
we focus on the need to define a stopping criterion. 
Due to expensive scanning time as well as the limited attentional capacity of subjects
it is imperative to devise formal stopping criteria. 
While there has been an extensive focus on online learning in the context of 
Bayesian optimization (e.g., minimizing cumulative regret \cite{Srinivas, Hoffman2}),
there are limited results relating to online stopping. 
In this work we propose two  stopping criteria and provide an empirical evaluation 
of their performance using data collected in the experiments of \cite{Lorenz}.

\section{Methods}

We begin by providing an overview of the study presented in Lorenz et al. \cite{Lorenz} in Section \ref{Overview} and a brief description of the Bayesian optimization technique in Section \ref{FixItSec}. The proposed stopping criteria are introduced and discussed in Section \ref{StopSection}.

\subsection{Overview of online study with fixed number of iterations}
\label{Overview}

\begin{figure}
 	\centering
      \includegraphics[scale=1.8]{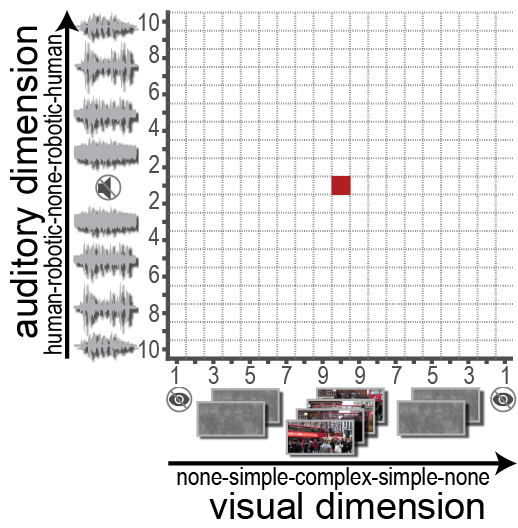}
  \caption{Experiment parameter space with 19 x 19 (361) possible combinations of audio-visual stimuli. For both, visual and auditory dimensions, stimuli varied in their perceptual complexity. The hypothezised optimal combination (derived from \cite{Braga}) for evoking the target brain state was the most complex visual stimulus in combination with no auditory input and was located in the center of the grid (red square). The objective of the Bayesian optimization algorithm was to traverse the parameter space and to learn the combination of audio-visual stimuli that best evokes the target brain state in a fully closed-loop manner.}
  \label{fig:grid}
\end{figure}

Five healthy subjects (2 females, mean age $\pm $ SD: 26.8 $\pm $ 3.6 years) participated in  the study of Lorenz et al. \cite{Lorenz}. Two target brain regions were identified based on previous literature \cite{Braga}: bilateral lateral occipital cortex and bilateral superior temporal cortex. The target brain pattern was defined as the maximum difference in brain activation between those regions (i.e., maximized occipital cortex activity with minimum superior temporal cortex activity). Visual and auditory stimuli were selected that have been shown to strongly activate those regions \cite{Braga}. Those stimuli were then systematically corrupted to span an experimental parameter space consisting of 361 different combinations of audio-visual stimuli (see Figure~\ref{fig:grid}). For both, visual and auditory dimensions, stimuli varied in their perceptual complexity from no visual/auditory input to video footage/a sentence spoken by a human voice in 19 discrete steps, respectively. In the online experiment, incoming whole-brain images (Siemens Verio 3T, TR = 2 sec) were pre-processed (i.e., motion-corrected and spatially smoothed) in real-time. Extracted timecourses of the target regions were further cleaned by removing low-frequency signal drifts as well as large signal spikes \cite{Koush}. After each audio-visual stimulus presented to the subject, the difference in brain level activation between the two regions was calculated in real-time using a general linear model approach. The result of this was then fed into the Bayesian optimization algorithm as loss function. See Lorenz et al. \cite{Lorenz} for further details.

\subsection{Bayesian optimization}
\label{FixItSec}

The underlying intuition behind the method of Lorenz et al. \cite{Lorenz}
is that the target brain pattern is a function 
of auditory and visual stimulus parameters. As such, the authors propose to learn 
the relationship
by modeling the observed brain activation as a sample from a Gaussian process (GP). 
This facilitates the use of a Bayesian 
optimization framework \cite{Snoek, BayesOptTutorial} to iteratively infer the combination of audio-visual stimuli which 
maximizes the target activation. 

Bayesian optimization algorithms are appealing due to their computational and mathematical simplicity. They are particularly relevant in the scenario of real-time fMRI 
as the aim is to optimize an unknown objective for which we do not have an analytical expression nor can we make formal statements regarding their properties (e.g., convexity); thus precluding traditional optimization techniques.



The Bayesian optimization algorithm proposed by Lorenz et al. \cite{Lorenz} is an iterative scheme where subjects are presented with a stimulus and their brain activation is measured. This activation is subsequently employed to update our belief regarding the relationship between the stimuli and target brain activation, captured in the posterior distribution of the GP \cite{GPbook}. The posterior distribution is subsequently employed to propose a new stimulus combination via the use of an acquisition function \cite{BayesOptTutorial}. The choice of acquisition function represents a trade-off between exploration of the stimuli space and exploitation. 
An advantage of employing a GP prior is that many acquisitions functions
depend solely on the prediction mean, $m(x)$, and 
variance function, $\sigma(x)$, of the model. 
Throughout this work, we follow \cite{Lorenz} and employ 
the expected improvement (EI) acquisition function \cite{BayesOptTutorial, Snoek}:
\begin{equation}
EI(x) = \left (m(x) - f(x^{+}) \right ) \Phi(z) + \sigma(x) \phi(z),
\end{equation}
where $\Phi$ and $\phi$ are the cumulative and probability density 
functions for a standard Gaussian random variable and
we define $z = \frac{m(x)-f(x^{+})}{\sigma(x)}$.
At each iteration, a new candidate is selected as $x_{new} = \underset{x}{\operatorname{argmax}} ~ \left \{ EI(x)\right \} $.


The use of a GP prior requires the specification of a mean and covariance function. 
As the objective of this work is to study the results of \cite{Lorenz}
in the context of stopping criteria, we select the same parameters as employed in 
the initial study.
Formally, 
a trivial zero-mean covariance, $m(x)=0$, was employed
together with a 
squared exponential covariance with automatic relevance discarding \cite{GPbook}. This required the specification of three hyper-parameters (two length-scale parameters and a noise parameter). 
The choice of these parameters is crucial to the success of Bayesian optimization methods and several sophisticated techniques have been proposed to select them \cite{BayesOptParameters}. In this work the hyper-parameters were selected via type-2 maximum likelihood on independent datasets \cite{GPbook, Lorenz}.


\subsection{Stopping criteria}
\label{StopSection}

Within the Bayesian optimization literature the focus has traditionally been on 
online learning \cite{Hoffman2}.
As a result, there is limited work studying stopping criteria which are particularly relevant in our scenario. In this section we propose two stopping criteria and study their performance empirically. The first stopping criterion is based on the Euclidean distance between consecutive proposed audio-visual stimuli combinations. The second approach exploits the wealth of information in the GP posterior to provide a more formal stopping criterion. This approach looks to establish a stopping criterion by combining two popular acquisition functions
as we discuss below.


\subsubsection{Euclidean distance}

As mentioned previously, the EI acquisition function \cite{BayesOptTutorial} is employed throughout
this work .
The EI acquisition function effectively looks to sample candidate stimuli that have a high posterior variance in the GP model, thereby favoring exploration in the face of uncertainty. This is in contrast to the probability of improvement (PI) acquisition  \cite{BayesOptTutorial} which is purely exploitative. The intuition behind this first stopping criterion is that when points within a similar region are consistently proposed, then it may be the case that the Bayesian optimization algorithm has entered a purely exploitative phase. 
While such a phase 
is desirable in many applications, the primary objective of Bayesian optimization within the context of real-time fMRI is to efficiently explore the set of potential stimuli and reliably learn
the relationship with observed brain activation. As such, once 
we enter a purely exploitative phase it may be the case that a sufficiently accurate model has been learnt.


Our proposed stopping criterion is as follows: at each iteration, we calculate the Euclidean distance (ED) between the point of the most recent observation and the proposed stimulus to be presented in the next iteration. 
Once this falls below a certain threshold, 
 the resulting Bayesian optimization algorithm is terminated.

A clear shortcoming of such a stopping criterion is that it cannot handle multimodal functions. In such a scenario it could be possible that each of the peaks is sampled iteratively, resulting in an experiment that is never terminated. Another shortcoming is the need to specify a threshold for the ED 
 which in practice is non-trivial and difficult to interpret. In order to
address these two issues we propose a second stopping criterion.

\subsubsection{Hybrid stopping criterion}
In order to address the shortcomings of the aforementioned approach
we propose a criterion based upon a hybrid of the EI and PI acquisition functions. 
Hybrids of acquisition functions have been studied in \cite{AcqPort}, however 
the motivation there was to improve sampling by proposing a portfolio of acquisition functions.
In contrast, our interest here is to combine two popular acquisition functions in order to obtain an intuitive and effective stopping criterion.


An advantage of both the PI and EI acquisition functions are that they are easily interpretable. In particular, the PI is a measure of the probability of observing an improvement in the latent objective function. Moreover, for a candidate stimuli, $x \in \mathbb{R}^{2}$, the PI acquisition function resembles a traditional $Z$-test  with the pivotal quantity 
\begin{equation}
z  =  \frac{m(x) - f(x^{+})}{\sigma(x)},  
\end{equation}
where $f(x^{+})$ is the maximum observed activation \cite{BayesOptTutorial}. 
As a result, calculating the PI is akin to calculating the $p$-value for the 
hypothesis test associated with the null $H_0: m(x) > f(x^+) $.

Based on 
this observation, we propose a stopping criteria by calculating the PI of each new candidate point, $x_{new}$, 
proposed by the EI acquisition function. 
Formally, we propose to terminate the Bayesian optimization algorithm when 
$PI(x_{new}) < \alpha$, where $\alpha$ is the significance level. 

The resulting stopping criterion requires one input parameter which can be interpreted as 
the probability of type 1 error (or of incorrectly terminating the algorithm).



\begin{wrapfigure}{r}{0.5\textwidth} 
 \begin{center}
      \includegraphics[scale=0.8]{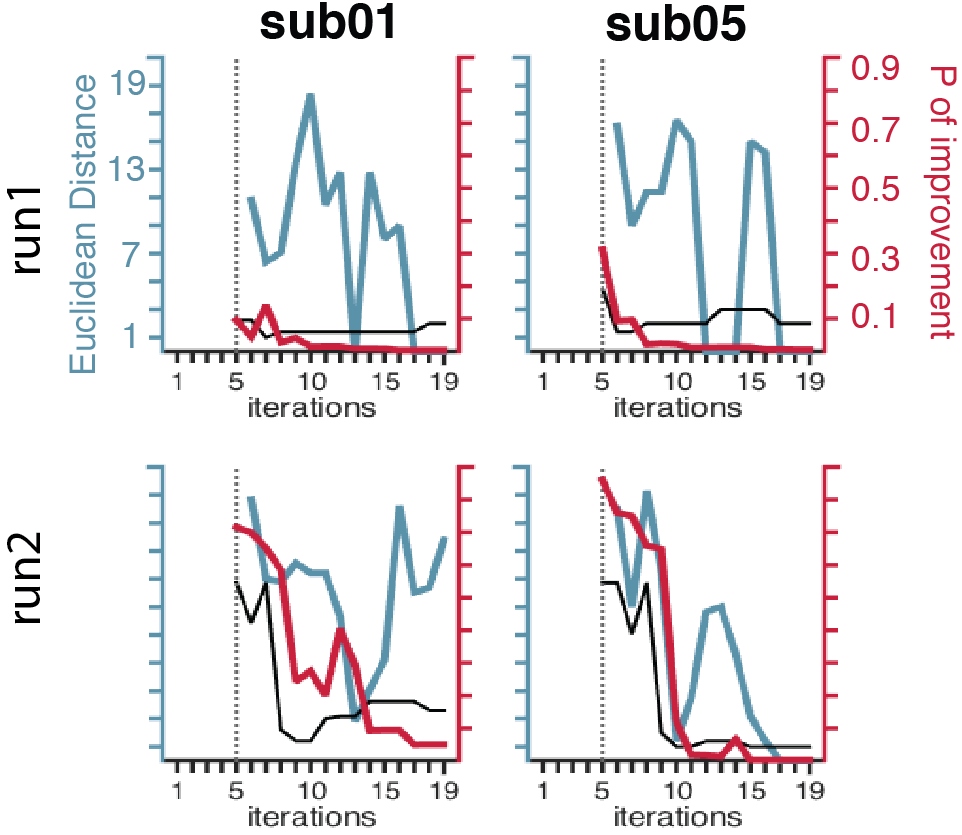}
      \end{center}
  \caption{The two proposed stopping criteria studied in two subjects on the first two initial runs. At each iteration, we computed the Euclidean distance (blue) and probability of improvement (red)  for each subject. As the first five iterations were used as a burn-in for a first estimate of the Bayesian model, they are not depicted here (grey dashed line). In order to infer accuracy of our model predictions we also computed the Euclidean distance between the predicted optimum and the hypothezised optimum at each iteration (thin black line).}
  \label{fig:results}
\end{wrapfigure}

\section{Results}
The performance of the proposed stopping criteria was studied empirically
using data from the original Bayesian optimization study \cite{Lorenz}. Data from the initial two runs for each of the five subjects was studied (sub04 only underwent one run). In the interest of clarity, we plot the proposed stopping criteria for two subjects in Figure~\ref{fig:results}.  We note that both proposed online stopping criteria seem suitable. In the majority 
of the runs, ED went to zero indicating that the same stimulus is proposed by the EI acquisition function over multiple iterations. For all of those runs, the PI at those points was consistently below $\alpha = 0.05$, 
suggesting that the experiment was run for a sufficiently long time.
In order to infer the accuracy of our model predictions we also computed the Euclidean distance between the predicted optimum and the hypothezised optimum at each iteration (thin black line in Figure~\ref{fig:results}). We found that accuracy would have not been impaired when stopping those runs earlier. In contrast, for three runs ED stayed above zero, indicating that for those runs a longer imaging acquisition might have been advantageous for making more accurate predictions. This can be inferred for example from run2 of sub01 in which the accuracy at the last iteration is worse (i.e., Euclidean distance between predicted and hypothesized optimum is higher) than for the first run.

\section{Discussion}

We present and empirically validate two stopping criteria for Bayesian optimization methods. The motivation behind this work was the use of such techniques
in conjunction with real-time fMRI methods to provide a more holistic understanding of how cognition and the brain inter-relate. Our work is closely related and complementary to the
recently proposed \textit{Automatic Neuroscientist} framework by Lorenz et al. \cite{Lorenz} where stopping criteria are required due to the expense of scanning and limited subject attention.

The first criterion proposed is simply based on the Euclidean distance between successive 
samples. While such an approach has clear shortcomings, it is shown to perform
relatively well in the audio-visual task studied in this work. We believe this may be due to the 
clear unimodal nature of the response function and we would expect such a stopping criterion to perform poorly
when faced with multimodal responses.

The second criterion we propose is based upon a hybrid of acquisition functions. In this manner, new candidate stimuli can be proposed via the EI acquisition function and
the Bayesian optimization algorithm is terminated when 
the PI of corresponding stimuli falls below a threshold $\alpha$. 
The advantage of such an approach is that $\alpha$ can be directly interpreted as the 
significance level in a traditional hypothesis test. 
An area for future work of such a stopping criteria would involve correcting for multiple comparisons, however
this is challenging when the number of comparisons is unknown a priori. 

Moreover, our results display clear 
inter-subject and inter-run variability. While for the majority of runs
scanning time could have been potentially reduced by up to seven observations (corresponding to approximately 2/3 minutes), for others, our stopping criteria indicated that further observations would have been required. Moreover, we find that employing the proposed stopping criteria would not affect the accuracy of the model predictions. 

 The \textit{Automatic Neuroscientist} has direct clinical relevance, since it can be extended to any desirable target brain pattern, including fMRI biomarkers relevant to clinical populations (e.g., a specific functional connectivity network configuration tracked in real-time \cite{Monti1, Monti2}). In this respect, the development of online stopping criteria is of paramount importance for the framework to be pulled through the translational pathway.  Various neurological and psychiatric disorders are characterized by impaired sustained attention, such as attention deficit disorder \cite{ADHD}, traumatic brain injury \cite{TBI}, or bipolar disorder \cite{bipolar}. In addition, children and the elderly only have limited attentional capacities.
 
Future work will require a formal analysis of the stopping criteria together 
with empirical validation on a more complex task.

\bibliographystyle{plainat}

\end{document}